\documentclass[envcountsame,envcountsect,oribibl]{llncs}
\RequirePackage{longtable}
\RequirePackage{amsmath}
\RequirePackage{amssymb}

\usepackage{graphicx}
\usepackage{placeins}
\usepackage{color}
\usepackage{cite}

\newcommand*{\EncE}[2]{E_{#1}(#2)}
\newcommand*{\match}[2]{#1 \approx_t #2}

\newcommand*{\E}[1]{[#1]_{K}}

\newcommand*{\andthe}{\wedge}

\title{\vspace*{-1.0cm}Client-Server Password Recovery \\ {\small (Extended Abstract)} \vspace*{-0.7cm}}

\author{{\L}ukasz Chmielewski\inst{1} \; Jaap-Henk Hoepman\inst{1,2} \; Peter van Rossum\inst{1} \vspace*{-0.2cm}} 

\institute{Digital Security Group\\
  Radboud University Nijmegen, the Netherlands\\ 
  \email{\{lukaszc,jhh,petervr\}@cs.ru.nl}
  \and
  TNO Information and Communication Technology, The Netherlands\\
  \email{jaap-henk.hoepman@tno.nl}\\
}

\newcommand{\ie}{{i.e.},\ }		% i.e.
\newtheorem{assumption}[theorem]{Assumption}

\newcommand{\keywords}[1]{\par\addvspace\baselineskip
\noindent\keywordname\enspace\ignorespaces#1}

\begin{document}

\bibliographystyle{plain}

\maketitle

%\footnotetext[1]{Jaap-Henk Hoepman is also with TNO, The Netherlands, e-mail: \newline \texttt{jaap-henk.hoepman@tno.nl}}

\vspace*{-0.8cm}\begin{abstract}
Human memory is not perfect -- people constantly memorize new facts and forget old ones. 
One example is forgetting a password, a common problem raised at IT help desks. 
We present several protocols that allow a user to automatically recover a password from a server using partial knowledge of the password. 
These protocols can be easily adapted to the \emph{personal entropy} setting \cite{schneier}, where a user can recover a password only if he can answer 
a large enough subset of personal questions.

\iffalse
%\looseness=-1
We introduce efficient password recovery methods that apply to two of the most common types of password based authentication systems. 
We call these methods 
we introduce \emph{client-server} password recovery, in which the recovery data should be stored at the server, and  
password recovery should be integrated into the login procedure. 
These methods apply to two of the most common types of password based authentication systems. 
We call these methods 
\fi
We introduce \emph{client-server} password recovery methods, in which the recovery data are stored at the server, and 
the recovery procedures are integrated into the login procedures. 
These methods apply to two of the most common types of password based authentication systems. 
The security of these solutions is significantly better than the security of presently proposed password recovery schemes. 
Our protocols are based on a variation of threshold encryption~\cite{gate, sharingdec, damgard03generalization} that may be of independent interest.
\end{abstract}

\vspace*{-0.7cm}\keywords{password recovery, threshold encryption scheme, private computing, personal entropy}

%\vspace*{-0.2cm}
\section{Introduction}

%\vspace*{-0.1cm}
People constantly memorize new facts, but also forget old ones. One quite
common example is forgetting a password. It is one of the most common problem raised
at IT help-desks. Therefore, many 
systems for password recovery (PR) have been built. The common aim of all these
systems is to provide reliable solutions for legitimate users 
%of the system 
to recover lost passwords or to receive a new password (i.e., resetting the old password), without significantly increasing the vulnerability against
attackers.

The simplest way to authenticate the user is to use an out-of-band channel, like a phone call, or show up physically at a system administrator. 
This is costly however, and cumbersome. 
More user-friendly, but less secure, is the common method used by many websites that store the password of the user in the clear and resend it to the user's email address on request. 
%new: 
Sometimes websites require a user to answer some personal question, like ``what is your mother's maiden name?''. 
However, this method is insecure because 
a password sent in cleartext can be easily intercepted and it is relatively easy to answer such a single question. 

Another widely used method to cope with forgetting passwords is a password reset system. 
In this system when a user forgets the password then the server sets a new password and emails the new password to the client 
(again maybe after answering a personal question). 
Now the legitimate user can regain system access easily. 
However, the security of this system depends heavily on the security of the email server, and therefore, this system is uninteresting from our point of view. 
%the end of new 

There is quite a lot of research on more sophisticated PR methods that
do not fully trust the server. One approach is to use secret sharing
\cite{AS79,grb79}. This solution divides a password into $n$ shares (that are
stored on trusted servers) in such a way that for the reconstruction, it is
necessary to collect at least a threshold $t$ of these shares. However, the
user still needs to authenticate somehow to the servers, and therefore this
system does not fully solve our problem.

In \cite{schneier} a PR system, based on \emph{personal entropy}, is proposed. 
In this system, a user is asked some questions about his personal history during password registration. 
The system generates a random secret key, and encrypts the real password with it. 
Subsequently, the answers given by the user are used to ``encrypt'' the random secret key. 
The user then stores the questions, the ``encryption'' of the secret value, and the encryption of the password on his computer. 
A secret sharing scheme is used to enable password recovery, even if some questions are answered incorrectly. 
% JHH: we can omit this
%
%The security of the system relies
%on the fact that the questions concern personal entropy that is hard to
%forget (e.g., ``what was the first car that you were driving?'') for the user,
%but it is hard to be guessed by an attacker. Therefore, this scheme
%\cite{schneier} strikes an attractive balance between convenience and security.
The drawback of this scheme is the lack of a rigorous security analysis.
In fact, \cite{BN00} demonstrates a serious weakness of this scheme: with the parameters recommended for a security level of $2^{112}$, 
the system is in fact vulnerable to an attack that requires only $2^{64}$ operations.

The ideas from \cite{schneier} were improved in \cite{nfaj2001}. 
This improved password recovery uses error-correcting codes instead of a secret sharing scheme. 
A rigorous security analysis is performed in the chosen model. 
%(we comment on this solution in Section~\ref{sec:fuzzy}). 
The solution of \cite{nfaj2001} uses techniques that are very close to secure sketches.
%and fuzzy extractors. 

%\looseness=-1 
Secure sketches and fuzzy extractors (described e.g., in~\cite{cryptoeprint}), and their robust versions~\cite{strong_extractor,bklr08},
are cryptographic tools useful for turning noisy information into cryptographic keys and securely authenticating biometric data. 
They may also be used to solve the password recovery problem. 
However, contrary to intuition, it seems hard to use these cryptographic primitives to solve password recovery in our most secure model, 
as show in Section~\ref{sec:fuzzy}. 

%The drawback of the system is a consequence of using error-correcting codes: for some given answers the adversary may recover the password successfully even if he has not got enough
%correct answers. 
%JHH: we can skip this
%Another related issue is hardening passwords using keystroke data, described in \cite{keystroke}.
%In this solution, the legitimate user's typing patterns (e.g., durations of keystrokes, and latencies between keystrokes) are combined with the user's password 
%to generate a hardened password that is convincingly more secure than conventional passwords against both online and offline attackers.

We believe that \cite{schneier, nfaj2001} 
%solutions 
are a significant step towards a practical PR solution. 
However, such so-called \emph{local} PR 
%password recovery 
systems are vulnerable to attackers that steal the recovery data from the user's machine 
(which is quite often inadequately secured) and then mount an \emph{offline} brute force attack to recover the password. 
To avoid this scenario, we introduce \emph{client-server} password recovery, in which the recovery data should be stored at the server, and PR 
%password recovery 
should be integrated into the login procedure. 
In such a setting (under the more reasonable assumption that the recovery data cannot be stolen from the secure server) an attacker can only 
perform an \emph{online} brute force attack. 
Security then can be increased by limiting the number of tries per account, or increasing the response time. 

\iffalse
Fuzzy extractors, described for example in \cite{cryptoeprint,bklr08} are useful in the context of password recovery as well.
Fuzzy extractors securely and efficiently turn noisy information into keys usable for cryptographic applications, in particular, 
securely authenticating biometric data. 
They also allow a user to authenticate to a server using a password that is `sufficiently close' to the correct password. 
%\todo{JHH: Can we say this?}
However, their security properties are similar to the error correcting approach
to password recovery outlined above, and therefore unsuitable as well.
\fi

Our contributions are the following. 
Firstly, we introduce the password recovery problem and the client-server PR security model, together with 
a short analysis of password authentication systems, in Section~\ref{sec:model}. 
All our client-server PR systems apply to a simple (low entropy) password login system. 
In all these PR systems, the client is stateless, and all recovery data is stored at the server. 
Our solutions reduce the entropy somewhat, but are still more secure than other approaches. 
Moreover, our ideas can be straightforwardly applied to the personal entropy system, as shown in Subsection~\ref{sec:exact}, 
making the recovery phase more secure.
We elaborate on using secure sketches and fuzzy extractors for PR in Section~\ref{sec:fuzzy}.
Subsequently, we present a new algorithm (Section~\ref{sec:local}) for local PR 
%(i.e., a solution that works in a similar setting as \cite{schneier, nfaj2001}), 
that is based on intraceability Assumption~$2$ from \cite{pinkas}. 
In Section~\ref{threshold}, 
we introduce a new variant of threshold encryption~\cite{gate, sharingdec, damgard03generalization}, 
called \emph{equivocal threshold encryption}, that does not provide validity proofs for the decryption shares. 
Combining these two, we present protocols for client-server PR 
%password recovery 
integrated into two classes of systems for password based login: the most common, hash based one in which the server keeps hashes of passwords 
but receives cleartext passwords during the login phase (Section~\ref{sec:rec2}), 
and the most secure solution, based on challenge response, in which the server never sees passwords in clear at all (Section~\ref{sec:rec3}). 
Moreover, in Appendix~\ref{sec:simple} we briefly present a simple substring-knowledge PR working in the challenge response setting. 
%new
Furthermore, all our password recovery systems can be easily modified to work as password reset systems. 
Due to space constraints we omit these easy transformations. 
%the end of new

%Password recovery where the server stores the passwords in the clear is trivial: the user can send the guess of the password, and the server can 
%apply any matching algorithm to verify that the guess is indeed close to the real password. 
Due to space constraints in this version of the paper, proofs of security and correctness of the presented protocols are short and informal. 

\iffalse
We note that our PR systems can be easily converted to the personal entropy system
But first we introduce the password recovery problem and the model in which it is solved. 

Our paper is structured as follows: we formally define the new PR model in Section~\ref{sec:model}, and introduce three PA schemes. 
The cryptographic primitives are presented in Section~\ref{sec:pre}.
Equivocable Threshold Encryption Scheme is shown in Section~\ref{threshold}. 
Later, in Section~\ref{sec:local} we present our local PR. 
In Sections \ref{sec:rec1}, \ref{sec:rec2}, and \ref{sec:rec3} we show our PR solutions. 
We finish our paper with conclusions in Section~\ref{sec:conclusions}.
\fi

%\vspace*{-0.3cm}
\section{Password Recovery Model}\label{sec:model}

%\vspace*{-0.1cm}
In this section we discuss the kinds of password authentication (PA) systems for which we consider password recovery, 
define exactly what we mean by password recovery, and talk about the kinds of adversaries our protocols need to withstand. 

\vspace*{-0.3cm}\subsection{Password Authentication (PA) Systems}

\begin{figure}[t!]
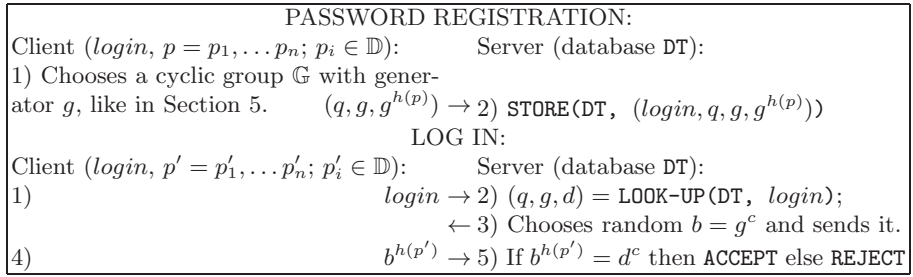

{\footnotesize
\begin{center}
\begin{tabular}{|p{5.7cm}p{0.3cm}p{5.7cm}|}
\hline
\multicolumn{3}{|c|}{PASSWORD REGISTRATION:}\\
Client ($login$, $p=p_1,\dots p_n$; $p_i \in \mathbb{D}$): & & Server (database \texttt{DT}): \\
1) Chooses a cyclic group $\mathbb{G}$ with generator $g$, like in Section~\ref{threshold}. 
\hfill $(q, g, g^{h(p)})$ & \strut\newline $\rightarrow$ & \strut\newline 2) \texttt{STORE(DT, $(login, q, g, g^{h(p)})$)}\\
%$
\multicolumn{3}{|c|}{LOG IN:}\\
Client ($login$, $p'=p'_1,\dots p'_n$; $p'_i \in \mathbb{D}$): & & Server (database \texttt{DT}): \\
1) \hfill $login$ & $\rightarrow$ & 2) $(q, g, d) =$ \texttt{LOOK-UP(DT, $login$)}; \\
 & $\leftarrow$ & 3) Chooses random $b = g^c$ and sends it.\\
4) \hfill $b^{h(p')}$ & $\rightarrow$ & 5) If $b^{h(p')}=d^c$ then \texttt{ACCEPT} else \texttt{REJECT}\\
\hline
\end{tabular}
\caption{challenge-response password authentication system\label{system2}}
\end{center}
}
\vspace*{-0.5cm}
\end{figure}

%In PA systems two kinds of participants are involved: users (also called clients) and servers. 
\looseness=-1
\vspace*{-0.1cm}Two kinds of participants are involved in PA systems: users (also called clients) and servers. 
Clients have a username (also called login) and a password \linebreak $p\text{=}p_1,\dots p_n$, where $p_i \in \mathbb{D}$ and $\mathbb{D}$ is 
the domain of characters of passwords ($\mathbb{D}$ is usually small, e.g., $|\mathbb{D}| \approx 100$).
For simplicity, we assume that clients always remember their logins, and that the length of the password $n$ is fixed for all users.

Initially, a client registers himself (in the \emph{registration phase}) with the server by submitting a username and an authenticator 
(derived from the password), which the server stores in its database.
Subsequently, the client can authenticate (in the \emph{log in phase}) to the server using his username and a proof of knowledge of the password. 
The server, using the authenticator from the database, and the proof of knowledge, can efficiently verify that the user knows the corresponding password. 
We distinguish three different PA schemes with respect to the security requirements. 
These systems differ in the way that authenticators and proofs of knowledge are defined: an authenticator can be equal to a password, a proof can 
be equal to a password (this is the case in hash based systems, where the server stores hashes of passwords), or neither of the above (which is the case for challenge-response type systems, an example of which is presented 
in Figure~\ref{system2}). 

The password recovery for the first system is trivial (because the server stores passwords in clear), and we omit it in this paper. 
The PR solutions for the other two PA systems are presented in Sections~\ref{sec:rec2} and~\ref{sec:rec3}, respectively.

\subsection{Client-Server Password Recovery (PR)}\label{sec:exact}

\vspace*{-0.1cm}A system for \emph{client-server} PR 
%password recovery 
has the same participants and log in routine as a PA system. 
Moreover, it provides an additional routine called \emph{password recovery} (PR), in which the client tries to recover the lost password. 
The password registration is also modified: besides submitting the login, and the authenticator, it also submits the recovery data. 
%(note, that the server should not be able to recover easily the password from the recovery data).
The client's input in the PR phase is login and a perturbed (incorrect) password $p'\text{=}p'_1,\dots p'_n$, while the server's input is the database 
with the logins and the registration data. 
Local password recovery is similar to client-server password recovery, except that the recovery data is stored locally at the client, and the 
recovery protocol is run locally at the client.

\looseness=-1
The requirement is that the client recovers the password, if and only if, $p'$ is similar to the password $p$ corresponding to his login. 
To be precise, we define similarity between strings $x$ and $y$ as $\match{x}{y}$ ($x$ matches $y$), if and only if, 
$t \le | \{ i \in \{1,\dots n\} : x_i = y_i \} |$. 
We assume that the parameters $n$ and $t$ are public.
%We also require in our most secure model that the server should learn from the PR run nothing more than from the PR data stored with the login. 
%Notice however, that a malicious server can always act as a client to himself, in order to recover the password by an exhaustive search using the PR routine. 

\looseness=-1
Note, that having partial knowledge of the password is a very similar recovery condition to the personal entropy one \cite{schneier, nfaj2001}. 
In the personal entropy system the client needs to answer some threshold of questions (i.e., $t$ out of $n$ questions) to recover the password.
The answers to the questions can be considered as an additional password, where every single answer can be treated as a letter. 
%Therefore, the systems \cite{schneier, nfaj2001} with $n$
%questions and threshold $t$ are nearly equivalent to a system in which a user
%can retrieve the password $p$ (of length $n$) if he knows more than $t$ letters
%of the auxiliary password. This auxiliary password can be even password $p$! 
%For simplicity, we consider in this paper that an auxiliary password is,
%indeed, $p$. 
It is easy to transform our systems to work with an auxiliary password, and therefore, with personal questions. 
We skip these straightforward transformations in this paper.

\looseness=-1
We develop our protocols based on the following assumptions.
%Firstly, 
We assume existence of the secure channels between the server and clients (which can be achieved using TLS connections). 
We work in the Random Oracle Model (ROM) \cite{random_oracle}, which means that we assume that hash functions work like random functions.
% JHH : we may assume that people know MAC's
%
Moreover, we use keyed hash functions, also called message authentication codes (MACs), 
of the form $\mathfrak{h}: \{0,1\}^{k} \times \mathbb{D} \rightarrow \mathbb{F}$, where $\mathbb{F}$ is a field. 
The first parameter of $\mathfrak{h}$ is a random string of length $k$ (the security parameter). For simplicity, we often omit this parameter in our descriptions.

We look for efficient protocols, \ie $O(n k)$, at the server side (because many clients might want to perform password recovery simultaneously), 
but we do allow a certain time penalty at the client side.

\vspace*{-0.3cm}\subsection{Adversaries and Security Requirements}

\iffalse
\vspace*{-0.1cm}This paper considers many kinds of adversaries. 
%In this section we describe the kinds of adversaries that our protocols need to defend against. 
%This paper considers many kinds of adversaries. 
Roughly speaking, we obey the following rule: 
adversaries impersonating clients are quite powerful, i.e., they are computationally bounded with respect to an arbitrarily large security parameter $k$, 
while the adversaries accessing the server database are computationally bounded by the size of the input data. 
Notice that, usually the size of the input is significantly smaller than $k$. 
\fi

%All adversaries are computationally bounded by a security parameter $k$, and cannot break our protocols only with an above negligible probability in $k$.
%Moreover, all adversaries may impersonate a client. 
All our client-server protocols defend against an adversary impersonating a client.
Such an adversary is computationally bounded by $k$ (but not by $n \log \mathbb{|D|}$) and is malicious \cite{OG04}, 
which means he can disobey the protocols routine. 
%Moreover, he can break our protocols only with a negligible probability in $k$.
%Such an adversary is malicious \cite{OG04}, which means it can disobey the protocols routine. 
This adversary tries to break a server's privacy that can be informally defined as follows. 
The impersonator, after any number of unsuccessful PR runs, can recover more information about the password, than 
following from the fact that the PR invocations failed, only with a negligible probability in $k$.
Notice however, that this adversary can always perform an online brute force attack on the PR routine (even using the password's distribution). 
But this is easily mitigated by adding timeouts or allowing only a fixed number of tries before blocking an account. 

We also consider an adversary accessing the server's database in all our client-server protocols. 
We model this adversary differently than the one impersonating client, because this adversary can perform offline brute force attack using the PR routine. 
Therefore, we define the adversary to not know the password distribution and to be computationally bounded with respect to $k$ and the parameters $n$, $t$, $|\mathbb{D}|$ 
(in a way that the problem from Assumption~\ref{ass:plr} is hard). 
The adversary tries to break a client's privacy that can be informally, defined as follows. 
For every two passwords $p'$ and $p''$, the corresponding two PR data instances are indistinguishable.
An adversary accessing local PR (see Section~\ref{sec:local}) is defined in the same way. 
%the end of new 

Only the challenge-response protocol (Section~\ref{sec:rec3}) is resistant against a fully corrupted server. 
The adversary corrupting the server is computationally bounded by $k$ and tries to gain information about client's password guesses 
from the data received in PR runs. 
We assume that this adversary is malicious in the sense, that he performs any actions to break the \emph{guesses} privacy. 
However, there is no point for him to alter the client's output: the client can easily verify correctness of the recovery by logging in. 
This approach is very similar to \emph{private computation} from \cite{pinkas}. 
The guesses privacy can be defined as follows: from a PR run the adversary gains negligible knowledge about the client's guess. 

\vspace*{-0.3cm}\section{Problems with Using Robust Fuzzy Extractors and Secure Sketches for Client-Server PR}\label{sec:fuzzy}

\vspace*{-0.1cm}
In this section we show the main problems of using secure sketches or fuzzy extractors solving client-server PR in our strongly secure model. 
Secure sketches and fuzzy extractors (see \cite{cryptoeprint}) can be used for turning noisy information into cryptographic keys and securely authenticating biometric data. 

\iffalse
Firstly, let's define secure sketches and fuzzy extractors. 
Let $\mathbb{F}$ be a field, $n~\in~\mathbb{N}$, and $\Delta$ a Hamming distance function in $\mathbb{F}^n$ 
(for $w,w' \in \mathbb{F}^n$ $\Delta(w,w')$ equals number of positions on which $w_i\text{=}w'_i$). 
An efficient $(\mathbb{F}^n, m, m', \tau)$-secure sketch is a pair of polynomial time randomized procedures, 
``sketch'' ($SS$) and ``recover'' ($Rec$), with the following properties. 
Firstly, $SS$ on input $w \in \mathbb{F}^n$ returns a bit string $s \in \{0, 1\}^*$. 
Secondly, the procedure $Rec$ takes an element $w' \in \mathbb{F}^n$ and a bit string $s \in \{0, 1\}^*$. 
The correctness property 
%of secure sketches 
guarantees that if $\Delta(w, w') \le \tau$, then $Rec(w', SS(w))\text{=}w$. 
If $\Delta(w, w') > \tau$, then no guarantee is provided about the output of $Rec$.
The security property guarantees that for any distribution $W$ over $\mathbb{F}^n$ with min-entropy $m$, 
the value of $W$ can be recovered by the adversary who observes $s$ with probability no greater than $2^{−m'}$. 
A robust version of secure sketch have the additional properties: (1) the phase $Rec(w', s)$ may fail (only) for $\Delta(w', w) > \tau$, 
and (2) a robust sketch detects when the adversary modifying $s$ (this is essential in the biometric authentication).
\fi
%Firstly, 
Now, let's define secure sketches and fuzzy extractors. 
Let $\mathbb{F}$ be a field, $n~\in~\mathbb{N}$, and $\Delta$ a Hamming distance function in $\mathbb{F}^n$. 
An $(\mathbb{F}^n, m, m', \tau)$-secure sketch is a pair of procedures, 
``sketch'' ($SS$) and ``recover'' ($Rec$), with the following properties. 
Firstly, $SS$ on input $w \in \mathbb{F}^n$ returns a bit string $s~\in~\{0, 1\}^*$. 
Secondly, the procedure $Rec$ takes an element $w' \in \mathbb{F}^n$ and a bit string $s~\in~\{0, 1\}^*$. 
The correctness property guarantees that if $\Delta(w, w') \le \tau$, then $Rec(w', SS(w))$ equals $w$. 
The security property guarantees that for any distribution $W$ over $\mathbb{F}^n$ with min-entropy $m$, 
the value of $W$ can be recovered by the adversary who observes $s$, with probability no greater than $2^{-m'}$. 
%A robust version of secure sketch have the additional properties: (1) the phase $Rec(w', s)$ may fail (only) for $dis(w', w) > \tau$, 
%and (2) a robust sketch detects when the adversary modifing $s$ (this is essential in the biometrical authentication).

\iffalse
An efficient $(\mathbb{F}^n, m, l, \tau,\epsilon)$-fuzzy extractor is a pair of polynomial time randomized procedures, 
``generate'' ($Gen$) and ``reproduce'' ($Rep$), with the following properties.
Firstly, the procedure $Gen$ on input $w \in \mathbb{F}^n$ outputs an extracted string $R \in \{0, 1\}^l$ and a helper string $P \in \{0, 1\}^*$.
Secondly, $Rep$ takes an element $w' \in \mathbb{F}^n$ and a string $P \in \{0, 1\}^*$ as inputs. 
The correctness property 
%of fuzzy extractors 
guarantees that if $\Delta(w, w') \le \tau$ and $P$ were generated by $(R,P)\text{=}Gen(w)$ then $Rep(w,P)\text{=}R$. 
If $\Delta(w, w') > \tau$, then there is no guarantee about the output of $Rep$. 
The security property guarantees that for any distribution $W$ over $\mathbb{F}^n$ with min-entropy $m$, 
the string $R$ is nearly uniform even for those who observe $P$: if $(R,P)\text{=}Gen(W)$, then statistical distance ${\bf SD} ((R,P), (U,P)) \le \epsilon$, 
where $U$ denotes uniform distribution of $|R|$ size string.
A robust version of fuzzy extractor have the additional properties: (1) the phase $Rep(w', P)$ may fail (only) for $\Delta(w', w) > \tau$, 
and (2) a robust sketch detects when the adversary modifying $P$ (this is essential in the biometric authentication).
\fi
An $(\mathbb{F}^n, m, l, \tau,\epsilon)$-fuzzy extractor is a pair of procedures, 
``generate'' ($Gen$) and ``reproduce'' ($Rep$), with the following properties.
Firstly, the procedure $Gen$ on input $w \in \mathbb{F}^n$ outputs an extracted string $R \in \{0, 1\}^l$ and a helper string $P \in \{0, 1\}^*$.
Secondly, $Rep$ takes an element $w' \in \mathbb{F}^n$ and a string $P \in \{0, 1\}^*$ as inputs. 
The correctness property 
%of fuzzy extractors 
guarantees that if $\Delta(w, w') \le \tau$ and $P$ were generated by $(R,P)\text{=}Gen(w)$ then $Rep(w',P)\text{=}R$. 
%If $dis(w, w') > \tau$, then there is no guarantee about the output of $Rep$. 
The security property guarantees that for any distribution $W$ over $\mathbb{F}^n$ with min-entropy $m$, 
the string $R$ is nearly uniform even for those who observe $P$.
%: if $(R,P)\text{=}Gen(W)$, then statistical distance ${\bf SD} ((R,P), (U,P)) \le \epsilon$, where $U$ denotes uniform distribution of $|R|$ size string.
A robust version of fuzzy extractor additionally detects whether the value $P$ got modified by an adversary 
(which is essential in the biometric authentication).
%have the additional properties: (1) the phase $Rep(w', P)$ may fail (only) for $dis(w', w) > \tau$, 
%and (2) a robust extractor detects when the adversary modifying $P$ (this is essential in the biometric authentication).

\looseness=-1 
Secure sketches 
%(standard or robust) 
can be used to solve local PR (Section~\ref{sec:local}) and client-server PR from Section~\ref{sec:rec2}.
Roughly speaking, the first case is close to the approach from \cite{nfaj2001}. Let's consider the second case. 
%Firstly, 
The client produces $s\text{=}SS(p)$ of his password $p$ and sends it to the server, who stores $s$. 
When the client invokes the PR routine by sending $p'$ then the server runs $p''\text{=}Rec(p', s)$ and if $\match{p'}{p''}$ then 
the server sends back $p''$. 
This solution is sound and secure, i.e, the server can guess $p$ with probability no greater than $2^{-m'}$. 
%The best for our purposes construction of secure sketches seems the one based on Reed-Solomon codes. 
%In this case and assuming that $m=n\log|\mathbb{D}|$ the following holds: $m'\text{=}(n - 2 \tau)\log|\mathbb{D}|$.
However, we do not see a way to transform this solution to the challenge response model, 
%(Section~\ref{sec:rec3}), 
because in this model the server is not allowed to see the password's guesses.
% during the recovery. 
%The main drawback of this solution is the fact that we do not see a way to transform this solution to challenge response model (Section~\ref{sec:rec3}). 
We leave finding the transformation of this solution to the challenge response model as a future work. 

%\looseness=-1 
It would appear that Robust Fuzzy Extractors (RFE) can be used to overcome this problem in, for example, the following way. 
First the client produces $(R,P)\text{=}Gen(p)$ and $E_{R}(p)$ (where $E$ is a symmetric encryption scheme, e.g., AES), and he sends $P$ and $E_{R}(p)$ to the server, who stores them. 
When the client invokes the PR routine, then the server sends the relevant $P, E_{R}(p)$ to the client. 
Now, the client can recover $R'\text{=}Rep(p',P)$, and try to decrypt: $Dec_{R'}(E_R(p))$. 
This solution is sound and seems secure. 
However, in our security model this protocol gives too much information to the adversary impersonating the client, 
because it allows an \underline{offline} dictionary attack. 
%To break the protocol the adversary needs to guess $l$ bits (notice, that practically always $l < m \le |\mathbb{D}|^n$).
We remind, that the adversary is computationally bounded by $k$ but not $n \log|\mathbb{D}|$.
Therefore, the adversary can simply guess $l$ bits (notice, that practically always $l < m \le n \log|\mathbb{D}|$), and break the protocol. 
Other solutions based on RFE seem to suffer to the same problem.

\vspace*{-0.3cm}\section{Local Password Recovery}\label{sec:local}

\vspace*{-0.1cm}As explained in the introduction, a client of \emph{local} password recovery, similarly to \cite{schneier, nfaj2001}, 
keeps the recovery data on his machine (there is no server). 
The client generates the recovery data and later on, tries to recover the lost password from the password guess and the recovery data. 
In Figure~\ref{fig:local} we present a solution for local PR. 
Its security is based on the following intraceability assumption derived from \cite{pinkas}, which is related to the polynomial list reconstruction problem.

\vspace*{-0.2cm}\subsubsection{The intraceability assumption.}

Let $C_{n,m}^{t,\alpha}$ denote the probability of distribution of sets generated in the following way: 
\begin{enumerate}
\item Pick a random polynomial $P$ over $\mathbb{F}$ (denote $|\mathbb{F}|=f$), of degree at most $t$, such that $P (0) = \alpha$.
\item Generate $nm$ random values $x_1,\dots x_{nm} \in \mathbb{F}$ subject to the constraint that all $x_i$ are distinct and different from $0$.
\item Choose a random subset $S$ of $n$ different indexes in $\{1,\dots nm\}$, and set $y_i = P(x_i)$ for all $i \in S$.
For every $i \not\in S$ set $y_i$ to be a random value in $\mathbb{F}$.
\item Partition the $nm$ $(x_i, y_i)$ pairs in $n$ random subsets subject to the following constraints. 
Firstly, the subsets are disjoint. Secondly, each subset contains exactly one pair whose index is in $S$ 
(hence $y_i = P (x_i)$) and exactly $m-1$ pairs whose indexes are not in $S$.
We denote these subsets as $S_i= \{(x_{(i,j)}, y_{(i,j)})\}$. 
%, where $1 \le j \le m$.
Output the resulting subsets. 
\end{enumerate}
The intractability assumption states that for any $\alpha, \alpha'$ the two probability ensembles $C_{n,m}^{t,\alpha}$, $C_{n,m}^{t,\alpha'}$ 
are computationally indistinguishable depending on the parameters $f$, $t$, $m$, and $n$.

\begin{assumption}[Assumption~2 from~\cite{pinkas}]\label{ass:plr}
Let $k$ be a security parameter, and let $n(k)$, $m(k)$, $t(k)$, $f(k)$ be at least linear polynomially bounded functions that define the parameters 
$n$, $m$, $t$ and $f$. 
Let $\mathcal{C}_{n,m}^{t,\alpha}$ and $\mathcal{C}_{n,m}^{t,\alpha'}$ be random variables that are chosen according to the distributions 
$C_{n,m}^{t,\alpha}$ and $C_{n,m}^{t,\alpha'}$, respectively. 
Then it holds that for every $\alpha, \alpha' \in \mathbb{F}$, the probability ensembles $\mathcal{C}_{n,m}^{t,\alpha}$ and $\mathcal{C}_{n,m}^{t,\alpha'}$ 
are computationally indistinguishable. 
\end{assumption}

\begin{figure}[t!]
\begin{center}
{\footnotesize
\framebox{
\begin{minipage}{11.75cm}
\textbf{Password Registration:}
The input is $p=p_1,\dots p_n$, where $p_i \in \mathbb{D}$, and $|\mathbb{D}|=m$. \\
The client: 
\vspace*{-0.2cm}\begin{enumerate}
\item Generates $v \in_R \{0,1\}^k$, and $n$ values $\{ \mathfrak{h}_{1}(p_1), \dots \mathfrak{h}_{n}(p_n) \}$. %We denote these values as $\tilde{x}_i$. 
Every $\mathfrak{h}_i$ is a MAC with implicit first parameter $v$ as described in Section~\ref{sec:exact}.
\item Generates $n$ random values $s_1, \dots s_n \in \mathbb{F}$ in such a way that points 
$\{(\mathfrak{h}_{1}(p_1), s_1), \dots (\mathfrak{h}_{n}(p_n), s_n)\}$ define a polynomial $P$ of degree $t-1$, and $P(0)$=$p$. 
\item Returns: $PR$=$(v,\{s_1-\mathfrak{g}_{1}(p_1), \dots s_n-\mathfrak{g}_{n}(p_n))\}$; each $\mathfrak{g_i}$ is a similar MAC to $h_i$. 
%Outputs: $PR$.
\end{enumerate}
\vspace*{-0.2cm}\textbf{Password Recovery:} The input is: $p'=p'_1,\dots p'_n$, $PR=(v, \{s'_1, \dots s'_n\})$.
%$PR=(v, \{s'_1 = s_1-g_{1}(p_1), \dots s'_n = s_n-g_{n}(p_n)\})$.
\vspace*{-0.2cm}\begin{enumerate}
\item The client computes set $S=\{(\mathfrak{h}_{1}(p'_1), s'_1+\mathfrak{g}_{1}(p'_1)),\dots (\mathfrak{h}_{n}(p'_n), s'_n+\mathfrak{g}_{n}(p'_n))\}$.
\item The client tries to reconstruct 
%the polynomial 
$P$ from any subset of $t$ elements of $S$ (that is $\binom{n}{t}$ checks). 
He checks whether for any potentially recovered polynomial $P'$ the following holds (let $p''\text{=}P'(0)$): $\match{p''}{p'}$
and $\{(\mathfrak{h}_{1}(p''_1), s'_1+\mathfrak{g}_{1}(p''_1)),\dots (\mathfrak{h}_{n}(p''_n), s'_n+\mathfrak{g}_{n}(p''_n))\}$ defines a polynomial of degree $n$. 
If it holds then he outputs $p''$.
If it does not hold for any $P'$ then the client outputs $\emptyset$.
\iffalse
He computes $p''=P(0)$ for these elements and checks whether $\match{p''}{p'}$, and $\{(\mathfrak{h}_{1}(p''_1), s'_1+\mathfrak{g}_{1}(p''_1)),\dots (\mathfrak{h}_{n}(p''_n), s'_n+\mathfrak{g}_{n}(p''_n))\}$ 
defines a polynomial of degree $n$. 
If it holds then he outputs $p''$.
If looking for the polynomial fails then he outputs $\emptyset$. 
This operations costs $O(\binom{n}{t})$ polynomial interpolations. 
\fi
\end{enumerate}
\end{minipage}
}}
\caption{Local Password Recovery \label{fig:local}}
\end{center}
\vspace*{-0.7cm}
\end{figure}

%%%%%%%%%%%%
In our applications the assumption's parameters are set as follows: $n$ and $t$ like in PR, $m = |\mathbb{D}|$ and $\mathbb{F} = \mathbb{Z}_q$, where $q$ is large prime. 
One may argue that $n$, $t$ and $|\mathbb{D}|$ are relatively small parameters 
(e.g., $n$ is the length of passwords) and that they might not deliver good security to the system. 
However, notice that in the personal entropy setting (i.e., the question-answer setting) 
%(the client needs to answer $t$ out of $n$ questions to recover the password) 
the parameters can be significantly enlarged. 
Moreover, we are not aware of any algorithm solving the assumption problem (i.e., finding $\alpha$) in our setting faster than by guessing $t$ proper points. 

We are conscious that for similar problems there exist fast solutions. 
For example, if in the above problem all $x_{(i,j)}=i$ then the problem can be solved fast (see \cite{BN00,boneh2002}). 
However, these fast algorithms do not solve the problem from Assumption~\ref{ass:plr}, as stated in~\cite{pinkas}. 
%%%%%%%%%%%%

\vspace*{-0.2cm}\subsubsection{The local PR solution.}
Now we describe the protocol. 
%from Figure~\ref{fig:local} 
%Firstly, 
In the first step the client prepares PR data: $v$ and $\{s_1-\mathfrak{g}_{1}(p_1), \dots s_n-\mathfrak{g}_{n}(p_n)\}$, such that 
$\{(\mathfrak{h}_{1}(p_1), s_1), \dots,(\mathfrak{h}_{n}(p_n), s_n)\}$ define a polynomial $P$ of degree $t-1$, for which $P(0)=p$.
Here, $\mathfrak{h}_{i}, \mathfrak{g}_{i}$ are hash functions (see Figure~\ref{fig:local}). 
Afterwards, the client forgets the password, and tries to recover it from 
$S=\{(\mathfrak{h}_{1}(p_1), s_1 -\mathfrak{g}_{1}(p_1) +\mathfrak{g}_{1}(p'_1)), \dots, (\mathfrak{h}_{n}(p_n), s_n-\mathfrak{g}_{n}(p_n) +\mathfrak{g}_{n}(p'_n))\}$.
If $\match{p}{p'}$ then he 
%the client 
obtains in $S$ at least $t$ proper points belonging to $P$, and can derive the password $P(0)$.
Otherwise, informally speaking, the client needs to solve the problem from 
%an instance of the intraceability 
Assumption~\ref{ass:plr}. 

\iffalse
\subsection{Correctness and Security}
The correctness of the recovery follows from the fact that if $\match{p}{p'}$ then the client has at least $t$ correct points 
(if $p_i=p'_i$ then $s_i-\mathfrak{g}_{i}(p_i)+\mathfrak{g}_{i}(p'_i)=s_i$), and can interpolate to get $p$. 
If $p' \not \approx_t p$ then $S$ does not contain enough points belonging to $P$ and the client's output is empty. 
\fi

\begin{theorem}[Local PR Security]\label{th:local}
An adversary $A$ attacking PR from Figure~\ref{fig:local} first produces two passwords $p_0,p_1$, and sends them to an oracle. 
Then the oracle chooses $b \in_R \{0,1\}$, performs password registration for $p_b$, and sends the result back. 
Finally, $A$ outputs his guess of $b$. 

$A$ succeeds with some probability $\frac{1}{2}+a$. We denote his advantage as $a$.
%Denote $A$'s advantage as $a$ for the probability of his success $\frac{1}{2}+a$. 
Working in ROM, no $A$ having non-negligible advantage 
exits under Assumption~\ref{ass:plr}.
\end{theorem}

%The proof sketch of this theorem is in the Appendix~\ref{ap:proof:local}.

\begin{proof}[sketch]
Assume to the contrary that there exists an adversary $A$, that attacks our local PR with non-negligible advantage. 
Using $A$, we construct an adversary $A^*$ that breaks Assumption~\ref{ass:plr}.
%, having an adversary $A$ that attacks our local PR.
Firstly, $A$ sends $p_0,p_1$ to $A^*$.
$A^*$ forwards them to an intraceability oracle (corresponding to Assumption~\ref{ass:plr}). 
This oracle chooses $b \in_R \{0,1\}$, and answers with $n$ subsets $S_i= \{(x_{(i,j)}, y_{(i,j)})\}$ sampled from $C_{n,|\mathbb{D}|}^{t,p_b}$. 
%(corresponding to  from assumption~\ref{ass:plr}), 
%where: $1 \le j \le m = |\mathbb{D}|$ and $P(0)= p_b$.
Now $A^*$ sends to $A$: $v \in_R \{0,1\}^{k}$, and $n$ random points in $\mathbb{F}$: $\{r_1, \dots r_n\}$. 
$A^*$ defines random oracles (representing $\mathfrak{h}_{i}$ and $\mathfrak{g}_{i}$) in the following way: for all $j \in \mathbb{D}$ and $i \in \{1,\dots n\}$:
%{\footnotesize \[
$RO_{\mathfrak{h}_i}(j)\text{=}x_{(i,j)} \text{ and } RO_{\mathfrak{g}_i}(j)\text{=}y_{(i,j)} - r_i$. \linebreak 
%\]}%
$A^*$ outputs the result of $A$. 
Notice, the importance of the implicit random parameter $v$, which lets random oracles, for two different PR runs, have different outputs 
(even for the same password).

Because of working in ROM, the distribution of $A$'s input, created in such a way by $A^*$ for $p_b$, is identical to the distribution of 
the client's input created in password registration (from Figure~\ref{fig:local}) for $p_b$. 
Therefore, $A^*$'s advantage is equal to $A$'s advantage, and Assumption~\ref{ass:plr} is broken. 
\qed
\end{proof}

%\section{Preliminaries}\label{sec:pre}

\iffalse
In this section we describe a few crypto primitives that we use in our solutions. 
One basic primitive that we us is symmetric encryption scheme (e.g., AES). 
In our proves of security we model symmetric encryption scheme as a finite pseudorandom function (FPRF) family 
(as it is described in \cite{bellare}). Roughly, we are assuming that as long as you don't know the underlying key, 
the input-output behavior of a cipher closely resembles that of a random function. 
We also assume that encryption process is randomized, so two encryption of the same message are indistinguishable. 
\fi

\vspace*{-0.3cm}\section{Equivocable Threshold Cryptosystem}\label{threshold}

\iffalse
In many applications, the ability to decrypt, i.e., the knowledge of the secret key, gives a huge power.
A classical way to reduce trust in such a secret's owner
%, and to increase the security, 
is to share the secret between many entities in such a way that cooperation between them is necessary to decrypt.
This property is obtained by threshold encryption (TE) schemes \cite{gate, sharingdec, damgard03generalization}. 
%Similarly, threshold signature scheme is introduced in \cite{shoup00practical}. 
The formal definition of TE scheme and an application of TE schemes for multi-party computation is in \cite{mpc-threshold}.
\fi

\vspace*{-0.1cm}In this section we define an equivocable threshold encryption (TE) scheme, and we present a slightly modified threshold ElGamal scheme 
(based on~\cite{gate}, and the ``normal'' ElGamal scheme \cite{ElGamal}) that is equivocable.
Subsequently, in Sections~\ref{sec:rec2} and~\ref{sec:rec3} we use this scheme to solve the PR problem. 

In \cite{sharingdec} a standard TE scheme consists of the following components. 
\emph{A key generation algorithm} $KG$ takes as input a security parameter $k$, the number of decryption servers $n$, the threshold 
parameter $t$ and randomness; it outputs a public key $pk$, a list $\alpha_1,\dots \alpha_n$ of private keys, and 
a list $vk_1,\dots vk_n$ of verification keys. 
\emph{An encryption algorithm} $Enc$ takes as input the public key $pk$, randomness and a plaintext $m$; it outputs a ciphertext $c$.
\emph{A share decryption algorithm} $SD$ takes as input the public key $pk$, an index $i \in \{1,\dots n\}$, the private key $\alpha_i$ and
a ciphertext $c$; it outputs a decryption share $c_i$ (called also partial decryption) and a proof of its validity $pr_i$.
Finally, \emph{a combining algorithm} $CM$ takes as input the public key $pk$, a ciphertext $c$, a list $c_1,\dots c_n$ of decryption shares,
a list $vk_1,\dots vk_n$ of verification keys, and a list $pr_1,\dots pr_n$ of validity proofs. 
It performs decryption using any subset of $\{c_1,\dots c_n\}$ of size $t$, 
%$\{c_{i_1},\dots c_{i_t}\}$ 
for which the corresponding proofs are verified. 
If there is no such set then $CM$ fails. 

An equivocable TE scheme consists of the same components as above, but: $KG$ does not produce 
verification keys, 
%(i.e., $vk_i$), 
$SD$ does not produce validity proofs, 
%(i.e., $pr_i$), 
and validity proofs are not part of $CM$'s input. 
Therefore, $CM$ simply checks if a decryption is possible for any subset $c_{i_1},\dots c_{i_t}$ (that is $\binom{n}{t}$ checks). 

A secure equivocable TE scheme should fulfill the standard TE security definition called threshold CPA \cite{sharingdec}. 
Notice, that omitting validity proofs does not help a malicious combiner to decrypt, because he possesses less data than for standard TE. 
A secure equivocable TE scheme moreover has the following properties. 
After any number of $CM$ invocations, a malicious combiner 
(which does not know any secret shares) gains no information about: 
(1) the plaintexts in unsuccessful runs (semantic security) and (2) the shares used in unsuccessful runs for producing partial decryptions. 
We formalize this intuition in Definition~\ref{equivocal}. 

\begin{definition}[Equivocable Security]\label{equivocal}
Define an oracle $O$. 
Firstly, $O$ performs algorithm $KG$ (for the parameters stated above). Then $O$ can be accessed by the following procedures:\\
$S(m)$; returns: an encryption $c$ of $m$, and correct decryption shares $c_1,\dots c_n$. \\
$I(m, i_1,\dots i_{t-1})$, where $i_1, \dots i_{t-1} \in \{1,\dots n\}$ and $|\{i_1, \dots i_{t-1}\}|=t-1$; produces an encryption $c$ of $m$, and 
$x_1, \dots x_n$, where $x_i = c_i = SD(pk,i,\alpha_i,c)$ if $i \in \{i_1, \dots i_{t-1}\}$, and $x_i=SD(pk,i,r_i,c)$ (where $r_i$ is a random value) otherwise;
returns $c,x_1,\dots x_n$.\\
$F(m)$; returns $c, SD(pk,1,r_1), \dots SD(pk,n,r_n,c)$; every $r_i$ is a random value.

First game (corresponds to property 1): 
\vspace*{-0.25cm}\begin{enumerate}
\item $O$ invokes $KG$, and sends a public key to a malicious combiner $C_1$. 
\item\label{step2} 
$C_1$ sends a message $m$ to the oracle $O$, which returns $S(m)$. 
This step is repeated as many times as the combiner wishes. 
\item $C_1$ chooses $m_0, m_1$ and sends them to the oracle.
\item\label{stepevery} 
$C_1$ chooses $i_1,\dots i_{t-1} \in \{1,\dots n\}$, and sends them to $O$, which chooses $b \in_R \{0,1\}$. 
Then $O$ sends back $I(m_b, i_1,\dots i_{t-1})$. 
This step is repeated as many times as the combiner wishes. 
\item $C_1$ repeats Step~\ref{step2}, and finally, outputs his guess of $b$. 
\end{enumerate}
\vspace*{-0.25cm}No polynomial time adversary $C_1$ guesses $b$ with a non-negligible advantage. 

Second game (corresponds to property 2): 
\vspace*{-0.25cm}\begin{enumerate}
\item $O$ invokes $KG$, and sends a public key to a malicious combiner $C_2$. 
\item\label{step22} The same like Step~\ref{step2} of $C_1$.
\item $C_2$ chooses $m$ and sends it to the oracle.
\item $C_2$ chooses $i_1,\dots i_{t-1} \in \{1,\dots n\}$, and sends them to $O$, which chooses $b \in_R \{0,1\}$. 
Then $O$ sends back $I(m,i_1,\dots i_{t-1})$ if $b=0$, and $F(m)$ otherwise.  
This step is repeated as many times as the combiner wishes. 
\item $C_2$ repeats Step~\ref{step22}, and finally, outputs his guess of $b$.
\end{enumerate}
\vspace*{-0.25cm}No polynomial time adversary $C_2$ guesses $b$ with a non-negligible advantage. 
\end{definition}

\iffalse 
The second game seem to easier to win than the first one, but in general it appears 
to be hard to prove. 
\fi

\vspace*{-0.3cm}\subsection{ElGamal Equivocable TE Scheme}\label{elgamal_eq}

\vspace*{-0.1cm}In this section we introduce our version of the ElGamal scheme and prove that this version is securely equivocable.

%\subsubsection{Discrete Log Setting.} 
Let $\mathbb{G}=<$\emph{g}$>$ denote a finite cyclic (multiplicative) group of prime order $q$ for which the Decision
Diffie-Hellman (DDH) problem is assumed to be infeasible: 
given $g^{\alpha}, g^\beta, g^\gamma$, where either $g^\gamma \in_R \mathbb{G}$ ($\in_R$ means that a value is chosen uniformly at random from a set) or 
$\alpha \beta = \gamma \mod q$, it is infeasible to decide whether $\alpha \beta = \gamma \mod q$. 
This implies that the computation Diffie-Hellman problem, which is to compute $g^{\alpha \beta}$ given $g^\alpha, g^\beta \in_R \mathbb{G}$, is infeasible as well. 
In turn, this implies that the Discrete Log problem, which is to compute $\log_g h = \alpha$ given $g^\alpha \in_R \mathbb{G}$, is infeasible.
We use the group $\mathbb{G}$ defined as the subgroup of quadratic residues modulo a prime $p$, where $q=(p-1)/2$ is also a large prime.
This group is believed to have the above properties. 

%\subsubsection{Homomorphic ElGamal Encryption.}
In the ElGamal scheme the public key consists of $q$, a generator $g$ of $\mathbb{G}$, and $h=g^\alpha$, while the private key is $\alpha \in \{0, \dots q-1\}$.
For this public key, a message $m \in \mathbb{G}$ is encrypted as a pair $(a, b) = (g^r , m h^r)$, with $r \in_R \mathbb{Z}_q$.
Encryption is multiplicatively homomorphic: given encryptions $(a, b)$, $(a', b')$ of messages $m, m'$, respectively,
an encryption of $m * m'$ is obtained as $(a, b) * (a', b') = (aa' , bb') = (g^{r+r'}, m*m'*h^{r+r'})$.
Given the private key $\alpha = \log_g h$, decryption of $(a, b) = (g^r , m h^r)$ is performed by calculating $b/a^\alpha = m$.

%\subsubsection{ElGamal Semantic Security.}
ElGamal semantic security can be defined using the following game. 
%Firstly, 
An oracle first sends $pk=(q,g,h)$ to an adversary. 
Then the adversary sends plaintexts $m_0, m_1 \in \mathbb{G}$ to the oracle, which answers, for $b \in_R \{0,1\}$, with $(g^r, m_b h^r)$. 
Finally, the adversary guesses $b$. 
The scheme is semantically secure if the adversary's advantage is negligible. 
%(i.e., the probability of his success is $\frac{1}{2}+\epsilon$, and $\epsilon$ is negligible). 
The ElGamal scheme achieves semantic security under the DDH assumption. 

%\subsubsection{Threshold ElGamal Decryption.}
In this paper we use a $(t, n)$-threshold ElGamal cryptosystem based on \cite{gate}, in which encryptions are computed using a public 
key $pk=(q,g,h)$, while decryptions are done using a joint protocol between $n$ parties. 
%$P_1,\dots, P_n$.
The $i$th party holds a share $\alpha_i \in \mathbb{Z}_q$ of the secret key $\alpha = log_g h$, where the corresponding 
$h_i = g^{\alpha_i}$ can be made public. As long as at least $t$ parties take part, decryption succeeds, whereas less than $t$ parties are
not able to decrypt.

We set the shares as follows: 
%Firstly, 
the dealer makes the polynomial \linebreak $f (x) = \sum_{i=0}^{t-1} a_i x^i \mod q$, by picking $a_i \in_R \mathbb{Z}_q$ (for $0 < i < t$) and $a_0=f(0)=\alpha$. 
In the original scheme, the $i$th share is $\alpha_i = f (i)$, while in our scheme $\alpha_i = f (x_i)$, and each $x_i \in_R \mathbb{Z}_q$ is made public.
The schemes security is based on linear secret sharing \cite{AS79}: $t$ points of a polynomial of degree
$t-1$ are sufficient to recover the polynomial and less points give no knowledge about $f(0)$.

The reconstruction of plaintext can be performed in the following way.
For some $c=(g^r, mh^r)$, it is required to have $t$ proper partial decryptions $g^{r \alpha_i}$ and $x_i$, which can be combined to compute (for any $x_0$): 
\vspace*{-0.25cm}\begin{equation}
{\footnotesize
g^{r f(x_0)} =  \prod_{i \in S} g^{r \alpha_i \lambda_{x_0,i}^S} \mod p \text{ where } \lambda_{x_0,i}^S = \prod_{i' \in S \backslash i} \frac{x_0-x_i}{x_i-x_{i'}} \in \mathbb{Z}_q \label{eq:int} 
}
\vspace*{-0.2cm}\end{equation}%
Hence, because $g^{r f(0)}$ can be computed, $c$ can be decrypted as follows: $mh^r/ g^{r \alpha} = m$. 
Equation~\ref{eq:int} describes a polynomial interpolation in the exponent.

%\subsubsection{ElGamal Equivocable TE Scheme.}
%This section shows 

We now show that our TE scheme is equivocable with respect to Definition~\ref{equivocal} under the DDH assumption.
For simplicity, we assume that the combiner receives only the data from unsuccessful invocations. 
However, the successful ones can be handled in a similar way to the security proof of \cite{gate}. 
We prove some lemmas, and then based on them we show that our scheme is equivocable. 

%+++++++++++++++++++
\begin{lemma}[Run Independence]\label{t:run1} 
We define the following game. 
Firstly, an adversary $A$ gets from an oracle a public key $pk=(q,g,g^\alpha)$, and parameters $t$, $n$. 
Secondly, the oracle: chooses $b \in_R \{0,1\}$, prepares a list of shares $\{ (x_1,\alpha_1), \dots (x_{n},\alpha_{n})\}$ with secret key $\alpha$, and sends 
$x_1, \dots x_n$ to $A$.
Then, $A$ chooses two plaintexts $p_0$ and $p_1$, and sends them to the oracle. 
Now, $A$ repeats as many times as he wishes the following step: 
$A$ chooses any $i_1,\dots i_{t-1} \in \{1,\dots n\}$ and sends them to an oracle, which returns:
\iffalse
\vspace*{-0.2cm}{\footnotesize\[\begin{array}{ccccc}
g^{r}, & p_b * g^{r \alpha}, & g^{r \alpha_{i_1}}, & \dots & g^{r \alpha_{i_{t-1}}}\\
\end{array}\vspace*{-0.2cm}\]}%
\fi
$g^{r}, p_b * g^{r \alpha}, g^{r \alpha_{i_1}}, \dots g^{r \alpha_{i_{t-1}}}$, \linebreak 
where $r \in_R \mathbb{Z}_q$ is chosen by the oracle.
Finally, $A$ outputs his guess of $b$. 

No polynomial adversary $A$ guesses $b$ with non-negligible advantage under the DDH assumption.
\end{lemma}

%The proof sketch of this lemma is in the Appendix~\ref{ap:proof0}. 

\begin{proof}[sketch]
Assume that $A$ asks the oracle for partial decryptions at most $d$ times (where $d$ is polynomial in $k$).
For simplicity, we assume here that $n=t=2$ and $d=2$. The proof for greater $n$, $t$, and $d$ can be made similarly.

Assume to the contrary that there exists an $A$, that wins the game with a non-negligible advantage $a$. 
Using $A$ we construct an adversary $A^*$ that breaks the ElGamal semantic security. 
Firstly, $A^*$ receives a public key $pk=(q,g,g^\alpha)$ from a ``semantic security'' oracle, and forwards it to $A$.
$A^*$ also generates $x_1, x_2 \in_R \mathbb{Z}_q$ and sends them to $A$. 
Then $A$ chooses plaintexts $p_0, p_1$, and sends them to $A^*$. 
Subsequently, $A^*$ forwards them to the oracle, which answers with $g^{r_1}, p_b g^{r_1 \alpha}$.
Now, $A^*$ chooses $j \in_R \{0,1\}$ and $\alpha_j \in_R \mathbb{Z}_q$. 
$A^*$ computes, using Equation~\ref{eq:int},
such $g^{\alpha_{j \oplus 1}}$  that points: $\{(0,\alpha), (x_1,\alpha_1), (x_2,\alpha_2) \}$ define a polynomial of degree $1$.
Then $A^*$ chooses $b' \in_R \{0,1\}$, and a random permutation $\pi: \{1,2\} \rightarrow \{1,2\}$.

Subsequently, $A$ asks for partial decryptions. 
When $A$ asks $e$th time ($1$st or $2$nd time) and $\pi(e)=1 \andthe i_1=j$ then $A^*$ answers:
$g^{r_1}, p_b * g^{r_1 \alpha}, g^{r_1 \alpha_j}$. 
\iffalse
{\footnotesize
\[\begin{array}{ccc}
g^{r_1}, & p_b * g^{r_1 \alpha}, & g^{r_1 \alpha_j} \\
\end{array}\]}%
\fi
If $\pi(e)=1$ and  $i_1 \not= j$ then $A^*$ halts and outputs a random bit.
Eventually, if $\pi(e)\not=1$ then $A^*$ sends to $A$ (for $r \in_R \mathbb{Z}_q$):
$g^{r}, p_{b'} * g^{r \alpha}, g^{r \alpha_2}$. 
\iffalse 
{\footnotesize\[\begin{array}{ccc}
g^{r}, & p_{b'} * g^{r \alpha}, & g^{r \alpha_2} \\
\end{array}\]}%
\fi 
Finally, $A^*$ returns $A$'s output.

Notice that in the case $\pi(e)=1$, the probability that $i_1 \not= j$ (and the attack stops with a random output) is $\frac{1}{2}$. 
Assume that it does not happen. 
Note, that if $b'=b$ then $A$'s input is well constructed and the probability that $A$ outputs $b$ is $\frac{1}{2} + a$. 
Otherwise, because of the random permutation $\pi$, $A$'s input is distributed independently of $b$ 
(even if the adversary asks less than $d=2$ times). Thus, the probability of $A$ guessing correctly is $\frac{1}{2}$ in this case. 
Therefore, the $A^*$'s advantage is $a / 4$.
\qed
\end{proof}

The proof for greater $n$ and $t$ is easy: $A^*$ can simply produce more data $\alpha_i$. 
In the case of $d > 2$, the proof is modified as follows. $A^*$ chooses randomly $t-1$ indexes and the corresponding shares. 
Then $A^*$ chooses $b' \in_R \{0,\dots d-1\}$, and constructs the answer to the $e$th question of $A$ ($1 \le e \le d$) 
as follows.
If $\pi(e)=1$ ($\pi$ is a random permutation of set $\{1,\dots d\}$) then, if $A^*$ knows $\alpha_{i_1},\dots \alpha_{i_{t-1}}$, 
then $A^*$ answers with 
$g^{r}, p_b * g^{r \alpha}, g^{r \alpha_{i_1}}, \dots g^{r \alpha_{i_{t-1}}}$. 
\iffalse
{\footnotesize 
\[\begin{array}{ccccc}
g^{r}, & p_b * g^{r \alpha}, & g^{r \alpha_{i_1}}, & \dots & g^{r \alpha_{i_{t-1}}}\\
\end{array}\]}%
\fi
If $\pi(e)=1$ and $A^*$ does not have corresponding shares then $A^*$ finishes and outputs a random bit.
Otherwise ($\pi(e)>1$), $A^*$ answers (using Equation~\ref{eq:int}) with: 
\vspace*{-0.2cm}{\footnotesize\[
g^{r}, p_x * g^{r \alpha}, g^{r \alpha_{i_1}},\dots g^{r \alpha_{i_{t-1}}}
\begin{cases}
x=0 & \text{ if } \pi(e)-1 \le b' \\
x=1 & \text{ otherwise } \\
\end{cases}
\vspace*{-0.2cm}\]}
Finally, $A$'s result is returned by $A^*$. 

This construction ensures that $A$'s input is either well constructed or, because of the permutation $\pi$, is produced independently of $b$. 
The probability of not returning a random bit (when $\pi(e)=1$) is $1/\binom{n}{t-1}$, and is non-negligible in $k$.
Details of this constructions are quite straightforward, and we omit them here. 

\begin{lemma}[Run Indistinguishability]\label{t:run2}
We define the following game. \linebreak Firstly, an adversary $A$ gets from an oracle a public key $pk=(q,g,g^\alpha)$, and parameters $t$, $n$. 
Secondly, the oracle: chooses $b \in_R \{0,1\}$, prepares a list of shares $\{ (x_1,\alpha_1), \dots (x_{n},\alpha_{n})\}$ with a secret key $\alpha$, 
and sends $x_1, \dots x_n$ to $A$. 
%The oracle also chooses $b \in_R \{0,1\}$.
Now, $A$ repeats as many times as he wishes the following step. 
$A$ chooses a set $I=\{i_1,\dots i_{t-1}\}$ (where each $i_f \in \{1,\dots n\}$ and $|I|=t-1$) and sends it to the oracle. \linebreak 
If $b=0$ then the oracle chooses $r \in_R \mathbb{Z}_q$ and answers with:
\iffalse
\vspace*{-0.2cm}{\footnotesize\[\begin{array}{ccccc}
g^{r}, & g^{r \alpha}, & g^{r \alpha_{i_1}}, & \dots & g^{r \alpha_{i_{t}}}\\
\end{array}\vspace*{-0.2cm}\]}%
\fi 
$g^{r}, g^{r \alpha}, g^{r \alpha_{i_1}}, \dots g^{r \alpha_{i_{t}}}$. 
Otherwise the oracle chooses $r, r_1, \dots r_{t-1} \in_R \mathbb{Z}_q$ and answers with: \linebreak 
\iffalse 
\vspace*{-0.2cm}{\footnotesize
\[\begin{array}{ccccc}
g^{r}, & g^{r \alpha}, & g^{r r_2}, & \dots & g^{r r_{t-1}}\\
\end{array}\vspace*{-0.2cm}\]}%
\fi 
$g^{r}, g^{r \alpha}, g^{r r_2}, \dots g^{r r_{t-1}}$. 
Finally, $A$ outputs his guess of $b$. 

No polynomial adversary $A$ guesses $b$ with non-negligible advantage under the DDH assumption.
\end{lemma}

The proof sketch of this lemma is in the Appendix~\ref{ap:proof1}. 

\begin{corollary}\label{col1}
We define the following game. 
Firstly, an oracle: chooses \linebreak $b~\in_R~\{0,1\}$, generates a public key $pk=(q,g,g^\alpha)$, 
and a list of random elements (in $\mathbb{Z}_q$): $\{ (x_1,\alpha_1), \dots (x_{n},\alpha_{l})\}$. 
Secondly, the oracle sends $l$, $pk$, and $x_1, \dots x_l$ to an adversary $A$. 
%The oracle also chooses $b \in_R \{0,1\}$.
The following action is repeated as many times as $A$ wishes:
if $b=0$ then the oracle chooses $r \in_R \mathbb{Z}_q$ and sends to $A$:
%{\footnotesize\[
$\begin{array}{ccccc}
g^{r}, & g^{r \alpha}, & g^{r \alpha_{1}}, & \dots & g^{r \alpha_{l}}\\
\end{array}$. 
%\]}%
Otherwise the oracle chooses $r, r_1, \dots r_{l} \in_R \mathbb{Z}_q$ and sends: 
%{\footnotesize\[
$\begin{array}{ccccc}
g^{r}, & g^{r \alpha}, & g^{r r_1}, & \dots & g^{r r_{l}}\\
\end{array}$. 
%\]}%
Finally, $A$ outputs his guess of $b$. 

No polynomial adversary $A$ that guesses $b$ with non-negligible advantage 
exists under the DDH assumption.
\end{corollary}
\begin{proof}
Follows directly from Lemma~\ref{t:run2} for parameters $t=l$ and $n=l+1$.
% and for the adversary always asking for $n-1$.
\qed
\end{proof}

%+++++++++++++++++++

Now based on Lemmas~\ref{t:run1},~\ref{t:run2}, we show that our TE scheme is equivocable.

\begin{theorem}[ElGamal Equivocable TE Scheme]\label{th:eqte}
The ElGamal TE \linebreak scheme described above in Section~\ref{elgamal_eq} is equivocable with respect to Definition~\ref{equivocal} under the DDH assumption.
\end{theorem}

\begin{proof}
%We assume (as mentioned above) that there are no 
Successful combining invocations can be handled like in the security proof from \cite{gate}. 
This theorem, for unsuccessful invocations, follows directly from Lemma~\ref{t:run1} for the first game, and from Lemma~\ref{t:run2} for the second game.  
%under the DDH assumption.
\qed
\end{proof}

%JHH
%
% This is too trivial
\iffalse
\section{Password Recovery when passwords are stored in clear}\label{sec:rec1}

If the server stores passwords in cleartext then the PR is trivial. 
During a PR invocation the client sends $login$ and perturbed password $p'$ to the server. 
The server, depending on whether $\match{p}{p'}$, might send the original password $p$ to the client.
This solution is efficient, but completely insecure from the client's point of view.
The solutions from next sections improves on this drawback.
\fi

\section{Password Recovery for the Hash based PA System}\label{sec:rec2}

In this section we present solutions that work for the most widely used PA system. 
%Firstly, 
We present first a simple and secure PR scheme, that has a functional drawback: the server's time 
complexity is too high for many scenarios. Secondly, we show the solution that eliminates this drawback. 

\vspace*{-0.3cm}\subsection{Simple PR System for the Hash based PA System}\label{simpleprs}

\vspace*{-0.1cm}In the simple PR system the server performs all important security actions. 
%: he checks whether the recovery is possible based on the client's guess. 
%, and then sends the result back to the client. 
%This solution combines local PR with the protocol from Section~\ref{sec:rec1}. 
During the registration the client sends to the server the login, and the password $p$. 
The server generates the local PR data, like in Section~\ref{sec:local}. 
Later, if the client wants to recover $p$, he sends a perturbed password $p'$ to the server, who 
runs the local PR routine (Section~\ref{sec:local}). 
If the recovery was successful then $p$ is sent to the client and the request is rejected otherwise. 
%If the server can recover the password from $PR$ and $p'$, then $p$ is sent to the client, and the request is rejected otherwise. 
The correctness and the security of this protocol follows directly from the corresponding local PR properties. 

%In this system
Notice, that the client's privacy is not protected during protocols run (the server even knows the result of PR). 
Furthermore, there are two significant drawbacks: $\binom{n}{t}$ checks on the server side, 
and we do not foresee any way to transform this protocol to work in the securer, challenge-response model. 
These problems are solved in Section \ref{complexprs}.

\vspace*{-0.3cm}\subsection{Improved PR System for the Hash based PA System}\label{complexprs}

\vspace*{-0.1cm}We improve the simple PR scheme by combining the equivocable TE scheme (Section~\ref{threshold}) with local PR. 
In this solution, the client checks whether the password recovery is possible. 
Therefore, the server's time complexity is efficient. 
The improved PR system is presented in Figure~\ref{scheme2}. 

%In the registration phase 
\looseness=-1
During registration the client first produces a public key $(q, g, g^\alpha)$ of the equivocable TE scheme, 
%(Section~\ref{threshold}), 
with the corresponding secret key $\alpha$ and computes an encryption $c$ of the password $p$.
Subsequently, he generates the PR data: secret values $v_1, v_2$ (they have the same meaning as $v$ in local PR) 
and points $\{(\mathfrak{h}_{i}(p_i), \alpha_i-\mathfrak{g}_{i}(p_i)) | i \in \{1,\dots n\} \}$. 
All the points $\{(\mathfrak{h}_{i}(p_i), \alpha_i)\}$ together with $(0,\alpha)$ define the polynomial of degree $t-1$. 
This construction is very similar to the local PR registration. 
The client also produces the login and the hash of the password for the PA system. 
Then all these data are stored on the server. 
Intuitively, the server cannot recover more than in local PR, because he stores 
the local PR data and an encryption of the password under the secret of the local PR data. 

If the client forgets the password then he invokes the PR routine by sending the login and a guess $p'$. 
Subsequently, the server produces, using the homomorphic property, a new encryption $c'$ of $p$. 
Afterwards, the potential partial decryptions $\{c'_i = {c'}^{y_i + \mathfrak{h}_{g}(p'_i)}| i \in \{1,\dots n\}\}$ are produced. 
Notice, that if $p'_i = p_i$ then ($\mathfrak{h}_{i}(p_i), c'_i$) is a proper partial decryption of $c'$. 
Later on, the server sends $v_1$ (so the client can compute $\mathfrak{h}$), $c'$, and $c'_1,\dots c'_n$. 
If $\match{p'}{p}$, then the client can easily obtain $p$, because he has at least $t$ proper decryptions. 
Otherwise, the client does not have enough correct decryptions to obtain $p$. 
Moreover, because of the equivocable property of the TE scheme, the client cannot recognize which partial decryptions are correct from the data 
from many unsuccessful PR runs. 

%The values 
$v_1$ and $v_2$ are implicit parameters for $\mathfrak{h}$ and $\mathfrak{g}$, respectively, that are used to make different local PR data indistinguishable. 
$v_1$ is public (it is send to the client before any authentication), while $v_2$ is not revealed to the client, so he cannot locally compute $\mathfrak{g}$. 

\begin{figure}[t!]
{\footnotesize
\begin{center}
\framebox{
\begin{minipage}{11.75cm}
\textbf{PASSWORD REGISTRATION:} The client's input is: $login$ and $p=p_1,\dots p_n$ ($p_i \in \mathbb{D}$); the server's input is his database.
\vspace*{-0.2cm}\begin{enumerate}
\item The client chooses $v_1, v_2 \in_R \{0,1\}^k$ and 
\item generates a public key of the $(t,n)$-TE scheme (Section~\ref{threshold}): $pk=(q, g, h$=$g^\alpha)$. 
Then he generates shares: $(x_1, \alpha_1), \dots (x_n, \alpha_n) \in {\mathbb{Z}_q}^2$ of the secret key $\alpha$, where 
$x_i=\mathfrak{h}_i(p_i)$. 
%, and all $\alpha_i$ are set like in Section~\ref{threshold}. 
$\mathfrak{h}$ is MAC (described in Section~\ref{sec:exact}) 
with implicit parameter $v_1$.
\item The client computes encryption of the password $p$: $c=(g^r, p * h^r)$, and 
\item produces $PR$=$(pk, v_1, v_2, c, \{\alpha_1-\mathfrak{g}_1(p_1), \dots \alpha_n-\mathfrak{g}_n(p_n)\})$; $\mathfrak{g}$ is MAC with implicit parameter $v_2$. 
Then he sends $(login, H(p), PR)$ ($H$ is from the PA system). 
\item The server stores $(login, H(p), PR)$ in his database.
\end{enumerate}
\vspace*{-0.2cm}\textbf{LOG IN:} The client sends his $login$, and $p$ to the the server, which accepts the client if $H(p)$ is equal 
to the corresponding value from the database.\\ 
\textbf{PASSWORD RECOVERY:} The client's input is: $login$ and $p'=p'_1,\dots p'_n$ ($p'_i \in \mathbb{D}$); the server's input is his database.
\vspace*{-0.2cm}\begin{enumerate}
\item The client sends $(login, p')$ to the server. 
\item The server performs:
  \begin{enumerate}
  \item finds $PR$=$(pk, v_1, v_2, c, \{y_1, \dots y_n\})$ corresponding to $login$ in the database. 
  \item\label{pr:rerand} re-randomizes $c=(a,b)$, by $c' = (a * g^{r'}, b * h^{r'})$. 
  %Note that $c'$ is an encryption of $p$.
  \item\label{pr:keys} produces $n$ potential partial decryptions of $c'$: $\forall_{i\in \{1,\dots n\}} c'_i = {a'}^{y_i + \mathfrak{g}_{i}(p'_i)}$.
  \item\label{pr:send} sends $v_1$, $pk$, $c'$, and the partial decryptions $\{c'_1,\dots c'_n\}$ to the client. 
  \end{enumerate}
\item\label{pr:ver} Using $\{(\mathfrak{h}_1(p_1),c'_1),\dots (\mathfrak{h}_n(p_n),c'_n)\}$, the client performs a $CM$ invocation from Section~\ref{threshold}. 
If a decryption $p''$ matches $p'$ then the client outputs $p''$.
\end{enumerate}
\end{minipage}
}
\caption{Improved PR for UNIX-based Log In \label{scheme2}}
\end{center}
}
\vspace*{-0.7cm}
\end{figure}

\vspace*{-0.4cm}\subsubsection{Correctness and Security.}

Correctness of the PR phase is straightforward: if $\match{p}{p'}$ then at least $t$ partial decryptions are correct and thus, 
the client can decrypt $c'$. 
Otherwise, the client does not have enough partial decryptions of $c'$. 
%and the output is empty. 

\iffalse
\begin{theorem}[The privacy of the client]\label{theorem:privacy_client}
If an adversary $A$, that has access to the PR data stored on the server, can distinguish (assuming ROM) between two different passwords of the client 
then either the DDH assumption does not hold, or Assumption~\ref{ass:plr} does not hold. 
\end{theorem}
\fi

\begin{theorem}[The privacy of the client]\label{theorem:privacy_client}
An adversary $A$ attacking the privacy of the client from Figure~\ref{scheme2} produces two passwords $p_0,p_1$, and sends them to an oracle. 
Then the oracle, chooses $b \in_R \{0,1\}$, performs the registration for $p_b$, and sends the result back. 
Finally, $A$ outputs his guess of $b$. 

%$A$ succeeds with some probability $\frac{1}{2}+a$. We denote his advantage as $a$.
%Denote $A$'s advantage as $a$ for the probability of his success $\frac{1}{2}+a$. 
Working in ROM, no $A$ having non-negligible advantage 
exits under the DDH assumption and Assumption~\ref{ass:plr}.
\end{theorem}

\begin{proof}[sketch]
Assuming that the DDH assumptions holds (and thus, the ElGamal is semantically secure), $A$ can break the scheme only by gaining the secret of the local PR data. 
Following Theorem~\ref{th:local}, if the local PR security is broken then Assumption~\ref{ass:plr} does not hold. 
\end{proof}

\begin{theorem}[The privacy of the server]\label{theorem:privacy_server}
Define an ideal situation to be one, in which an adversary tries PR by sending his guess $p'$ of the password $p$ to the server, 
who returns $p$ if $\match{p'}{p}$, and the empty string otherwise. 
Now, define a simulator as an algorithm that works in the ideal situation, and acts as a server to an adversary $A$ attacking the privacy of the server. 

In ROM and under the DDH assumption, there exists a simulator $I$ such that no adversary $A$ can distinguish between $I$ and the real server 
(from Figure~\ref{scheme2}) with non-negligible advantage. 
\end{theorem}

The proof sketch of this lemma is in the Appendix~\ref{ap:proof2}. 

\vspace*{-0.4cm}\subsubsection{Complexity.}

During the registration the client sends a public key, two secret values (of length $k$), the login, the hash of the password, an encryption of the password, and $n$ perturbed shares. 
The complexity of this phase can be bound by $O(nk)$ bits. 
In the PR phase the server sends the public key, an encryption of password, and $n$ potential partial decryptions. 
This totals to $O(nk)$ bits. 

The registration is performed efficiently by the participants. In the PR phase the server's performance is fast (main load is $n$ exponentiations), 
while the client's time complexity involves $\binom{n}{t}$ polynomial interpolations (Step~\ref{pr:ver}). 
%However, this is not a drawback because the client does not often perform PR.
% hence his time complexity is not a requirement of the system.

\iffalse
This protocol is efficient and fulfill the requirements of classical ``hash'' PA systems. The PR phase can be easily added to existing systems. 
This solution can be also easily modified to work in the personal entropy setting. 
\fi

\section{Password Recovery for the Challenge-Response System}\label{sec:rec3}

In this section we present a PR solution for challenge response login system, where the password or the guess of the password is never sent to the server. 
%the strongest PR security requirements.  
We combine the protocol from Section~\ref{complexprs} with $OT_l^n$ oblivious transfer (see below). 
The challenge-response PR protocol is shown in Figure~\ref{scheme3}.

There are two participants in the OT protocol: Receiver, who wants to obtain some information from a remote database and Sender that owns the database. 
OT can be formalized as follows. 
During a $2$-party $1$-out-of-$n$ OT protocol for $l$-bit strings ($OT_l^n$), Receiver fetches $S[q]$ from the Sender's database 
$S = (S[1],\dots S[n])$, $S[j] \in \{0, 1\}^l$, so that a computationally bounded Sender does not know which entry Receiver is learning. 
Moreover, we assume information-theoretically privacy of Sender (it means that Receiver obtains only desired $S[q]$ and nothing more). 
Such $OT_l^n$ scheme is presented in \cite{Lipmaa05}. This OT protocol works in bit communication $O(k \log^2 n + l \log n)$, 
low degree polylogarithmic Receiver's time computation and linear time Sender's computation. 
This is the fastest oblivious transfer protocol to the best of our knowledge. 

%\looseness=-1
This system is very similar to the one from Section~\ref{complexprs}.  
However, the log in routine is different (i.e., the challenge-response one is used), and the PR routine is a bit modified. 
The client does not send the guess $p'\text{=}p'_1,\dots p'_n$ directly to the server. 
Instead, he obtains partial decryptions corresponding to $p'$ in an oblivious way, as follows. 
For each $i \in \{1,\dots n\}$, the server prepares a potential partial decryption $c'_i$ for all possible $|\mathrm{D}|$ letters (Step~\ref{step:ot}).
%The server prepares potential partial decryptions for all possible $nm$ passwords. 
Then the client asks for partial decryptions for guess $p'\text{=}p'_1,\dots p'_n$ by performing oblivious transfer $n$ times: for every letter $p'_i$ separately. 
In this way, the server does not gain information about $p'$, and the client cannot ask for more than one partial decryption per OT protocol. 
The protocol's security follows from the security of OT and the security properties of the scheme from Section~\ref{complexprs}. 

\begin{figure}[t!]
{\footnotesize
\begin{center}
\framebox{
\begin{minipage}{11.75cm}
\textbf{PASSWORD REG.:} like in Fig.~\ref{scheme2}, but instead of $H(p)$, values $g,g^{H(p)}$ are sent. \\ 
\textbf{LOGGING IN:} like in the challenge-response PA system (Figure~\ref{system2}). \\ 
\textbf{PASSWORD RECOVERY:} The client's input is: $login$ and $p=p'_1,\dots p'_n$; $p'_i \in \mathbb{D}$; the server's input is the database.
\vspace*{-0.2cm}\begin{enumerate}
\item The client sends $(login, p')$ to the server. 
\item The server, using $login$, finds $PR$=$(pk, v_1, v_2, c, \{y_1, \dots y_n\})$ in the database. 
Then he re-randomizes $c=(a,b)$: $c' = (a * g^{r'}, b * h^{r'})$  
%Note that $c'$ is a valid encryption of $p$.
and sends $v_1$, $pk$, $c'$.
%, to the client.   
\item\label{step:ot} For $i \in \{1,\dots n\}$, the client and the server performs $OT_m^{b}$ protocol, 
where $|\mathbb{D}|$=$m$ and $b$ is a partial decryption's bit size. 
%of the ciphertext in the following manner.
The server acts as Sender with the database: 
{\footnotesize 
\[
S[j]={c'}^{y_i + \mathfrak{g}_i(j)}, \text{ for all } j \in \mathbb{D}
\]}%
and the client acts as Receiver with index $q=p_i$. The client's output is $S[q]$. 
\item\label{pr2:ver} The same like Step~\ref{pr:ver} in PR from Figure~\ref{scheme2}. 
%The client tries to recover the password in the same way like in protocol from Figure~\ref{scheme2} (cause he gets actually 
%the same kind of data). It is $\binom{n}{t}$ polynomial interpolations. 
\end{enumerate}
\end{minipage}
}
\caption{challenge-response PR\label{scheme3}}
\end{center}
}
\vspace*{-0.7cm}
\end{figure}

\vspace*{-0.3cm}\subsection{Correctness and Security}

\vspace*{-0.1cm}We give an informal intuition about the theorems and the proofs. 
%The formalization of is straightforward, and is omitted due to space constraints. 
The proof of the correctness and the privacy of the client outside the protocol runs are the same as for the system from Figure~\ref{scheme2}. 
The proof of the privacy of the server is the same as the one for PR from Figure~\ref{scheme2}, assuming that the OT is secure. 
%Additional argument in this case is that the client does not obtain more potential decryption shares, cause of the security properties of oblivious transfer 
%(Section~\ref{sec:oblivious}). 
The privacy of the client during PR runs is maintained by using OT (the server cannot gain any information about the client guess $p'_1,\dots p'_n$).

\vspace*{-0.3cm}\subsection{Complexity}

\vspace*{-0.1cm}Only the PR phase is significantly different from the system from Figure~\ref{scheme2}. 
%In the PR phase the server sends the public key and an encryption of the password (this is $O(k)$ bits). 
%Subsequently, the server and the client performs  
The major payload comes from $n$ runs of $OT_{|\mathbb{D}|}^{O(k)}$ protocols.  
This can be bound by $O(n (k \log^2 |\mathbb{D}| + k  \log |\mathbb{D}|)) = O(n k \log^2 |\mathbb{D}|)$ bits. 
The bit complexity of this PR, although greater than the one from Figure~\ref{scheme2}, is still efficient.  
%This is more than for Protocol from Figure\ref{scheme2}, but still relatively efficient. 

In the PR protocol the time complexity of the client is relatively high and follows from $\binom{n}{t}$ polynomial interpolations. 
%(but as explained before, it is not a serious drawback).
The main drawback of this protocol is the time complexity of the server, who
acts as Sender in OT, using $O(n*|\mathbb{D}|)$ operations.  However, for the 
relatively small domain of letters $\mathbb{D}$, and due to the fact that PR is
performed rarely, this solution is still quite feasible.  This drawback might
be of greater impact if we use this protocol in the personal entropy setting (i.e., the question-answer setting), 
where $|\mathbb{D}|$ might be larger.

%This protocol is efficient and obey security requirements of challenge-repons logging in systems. Password recovery phase can 
%be also easily added to existing systems. Furthermore the security parameters $t$ and $n$, and even $D$ can be made client specific. 

\section{Conclusions}\label{sec:conclusions}

In this paper we have presented secure and efficient solutions for password
recovery, where the recovery data is stored securely at the server side.
Our solutions apply to all common types of password authentication systems,
without significantly lowering their security. We have introduced a variant of
threshold encryption, called equivocable, that serves as a building block to
our solutions, and that may be of independent interest as well.

Further research could be aimed at alternative definitions of password similarity, that also include reordering of password letters (which is a common mistake). 
Other issues that can be improved are the 
$\binom{n}{t}$ time complexity at the client side, and  
the server's time complexity in the challenge-response protocol (Section~\ref{sec:rec3}). 
%Also, it seems that an $\binom{n}{t}$ time complexity at the client side is unavoidable, but no proof or counterexample is known as of date.

\bibliography{passwordrecovery}

\begin{thebibliography}{10}

\bibitem{random_oracle}
Mihir Bellare and Phillip Rogaway.
\newblock Random oracles are practical: a paradigm for designing efficient
  protocols.
\newblock In {\em CCS '93: Proceedings of the 1st ACM conference on Computer
  and communications security}, pages 62--73, New York, NY, USA, 1993. ACM.

\bibitem{grb79}
G.R. Blakley.
\newblock Safeguarding cryptographic keys.
\newblock In {\em AFIPS Conference Proceedings}, volume~48, pages 313--317,
  June 1979.

\bibitem{BN00}
Daniel Bleichenbacher and Phong~Q. Nguyen.
\newblock Noisy polynomial interpolation and noisy chinese remaindering.
\newblock In {\em EUROCRYPT}, pages 53--69, 2000.

\bibitem{boneh2002}
Dan Boneh.
\newblock Finding smooth integers in short intervals using crt decoding.
\newblock {\em J. Comput. Syst. Sci.}, 64(4):768--784, 2002.

\bibitem{damgard03generalization}
Ivan Damgard, M.~Jurik, and J.~Nielsen.
\newblock A generalization of paillier's public-key system with applications to
  electronic voting, 2003.

\bibitem{cryptoeprint}
Yevgeniy Dodis, Rafail Ostrovsky, Leonid Reyzin, and Adam Smith.
\newblock Fuzzy extractors: How to generate strong keys from biometrics and
  other noisy data.
\newblock Cryptology ePrint Archive, Report 2003/235, 2003.
\newblock \url{http://eprint.iacr.org/}.

\bibitem{schneier}
Carl Ellison, Chris Hall, Randy Milbert, and Bruce Schneier.
\newblock Protecting secret keys with personal entropy.
\newblock {\em Future Generation Computer Systems}, 16(4):311--318, 2000.

\bibitem{sharingdec}
Pierre-Alain Fouque, Guillaume Poupard, and Jacques Stern.
\newblock Sharing decryption in the context of voting or lotteries.
\newblock In {\em FC '00: Proceedings of the 4th International Conference on
  Financial Cryptography}, pages 90--104, 2001.

\bibitem{nfaj2001}
Niklas Frykholm and Ari Juels.
\newblock Error-tolerant password recovery.
\newblock In {\em CCS '01: Proceedings of the 8th ACM conference on Computer
  and Communications Security}, pages 1--9, New York, NY, USA, 2001. ACM.

\bibitem{ElGamal}
Taher~El Gamal.
\newblock A public key cryptosystem and a signature scheme based on discrete
  logarithms.
\newblock In {\em Proceedings of CRYPTO 84 on Advances in cryptology}, pages
  10--18, New York, NY, USA, 1985. Springer-Verlag New York, Inc.

\bibitem{OG04}
Oded Goldreich.
\newblock {\em Foundations of Cryptography: Volume 2, Basic Applications}.
\newblock Cambridge University Press, New York, NY, USA, 2004.

\bibitem{bklr08}
Bhavana Kanukurthi and Leonid Reyzin.
\newblock An improved robust fuzzy extractor.
\newblock In {\em SCN}, pages 156--171, 2008.

\bibitem{Lipmaa05}
Helger Lipmaa.
\newblock An oblivious transfer protocol with log-squared communication.
\newblock In Jianying Zhou, Javier Lopez, Robert~H. Deng, and Feng Bao,
  editors, {\em ISC}, volume 3650 of {\em Lecture Notes in Computer Science},
  pages 314--328. Springer, 2005.

\bibitem{pinkas}
Moni Naor and Benny Pinkas.
\newblock Oblivious polynomial evaluation.
\newblock {\em SIAM J. Comput.}, 35(5):1254--1281, 2006.

\bibitem{strong_extractor}
Naom Nisan and Amnon Ta-Shma.
\newblock Extracting randomness: a survey and new constructions.
\newblock {\em J. Comput. Syst. Sci.}, 58(1):148--173, 1999.

\bibitem{PP99}
Pascal Paillier.
\newblock Public-key cryptosystems based on composite degree residuosity
  classes.
\newblock In {\em Advances in Cryptology -- EUROCRYPT}, pages 223--238, May
  1999.

\bibitem{gate}
Berry Schoenmakers and Pim Tuyls.
\newblock Practical two-party computation based on the conditional gate.
\newblock In Pil~Joong Lee, editor, {\em ASIACRYPT}, volume 3329 of {\em
  Lecture Notes in Computer Science}, pages 119--136. Springer, 2004.

\bibitem{AS79}
Adi Shamir.
\newblock How to share a secret.
\newblock In {\em Communications of the ACM, vol. 22, n.11}, pages 612--613,
  November 1979.

\end{thebibliography}

\appendix

\section{Simple Substring-Knowledge Password Recovery in the Challenge-Response Setting}\label{sec:simple}

In this appendix we present a simple and efficient substring-knowledge challenge-response PR scheme that uses an additively homomorphic encryption scheme. 
In order for a client to recover a password it needs to prove to the server that he remembers a substring of the original password.

Let $\E{\cdot}$ denote a homomorphic encryption function with a public key $K$. The homomorphic cryptosystem supports the following two 
operations, which can be performed without knowledge of the private key. Firstly, given the encryptions $\E{a}$ and $\E{b}$ of $a$ and $b$, 
one can efficiently compute the encryption of $a + b$, denoted $\E{a + b} := \E{a} +_h \E{b}$. Secondly, given a constant $c$ and the encryption $\E{a}$ of $a$, 
one can efficiently compute the encryption of $c \cdot a$, denoted $\E{a \cdot c} := \E{a} \cdot_h c$. 
These properties hold for suitable operations $+_h$ and $\cdot_h$ defined over the range of encryption function. 
An example of such an encryption scheme is Paillier's cryptosystem~\cite{PP99}.
%, where operation $+_h$ is a multiplication and $\cdot_h$ is an exponentiation. 

In the registration phase the client sends, besides data necessary for logging in, $h_1(p_{1,t}), h_2(p_{2,t+1}), h_{n-t+1}(p_{n-t+1,n})$ 
(for simplicity, we denote $p_{i,w}= p_i,\dots p_w$) and $\EncE{H_1(p_{1,t})}{p}, \dots \EncE{H_{n-t+1}(p_{n-t+1,n})}{p}$, 
where $\EncE{sk}{.}$ is a symmetric encryption scheme and $H_i, h_i$ are hash functions. 
Notice, that to recover the password $p$, it is necessary to derive some $h_i(p_i,\dots p_{i+t-1})$ or $H_i(p_i,\dots p_{i+t-1})$, 
and it (assuming ROM) is only possible by obtaining any substring $p_i,\dots p_{i+t-1}$. 

Later on, in the PR phase, the client produces a public key $K$ of the homomorphic encryption scheme, and sends it to the server together with \linebreak
$\E{h_1(p'_{1,t})}, \dots \E{h_{n-t+1}(p_{n-t+1,n})}$. 
Then the server computes: %\linebreak
$\{ \E{ (h_{i}(p'_{i,i+t-1})-h_i(p_{i,i+t-1})) * r_i + \EncE{H_i(p_{i,i+t-1})}{p}} | i \in \{1,\dots n-t+1 \} \}$,
(where $r_i$ are random values), and sends this set to the client. 
The client decrypts the values from the received set 
and checks if he can decrypt these values with any $H_1(p_{i,t}), H_2(p_{2,t+1}), H_{n-t+1}(p_{n-t+1,n})$ (then he derives $p$). 

The scheme is correct, since if $h_i(p'_{i,i+t-1})=h_i(p_{i,i+t-1})$ then the client obtains 
$\EncE{H_i(p_{i,i+t-1})}{p}$, and he can easily decrypt it. Otherwise, the value received is random (because $r_i$ are random) and therefore, the client 
cannot successfully decrypt. The privacy is protected by the security of the encryption schemes.

\section{Proof Sketch of Lemma~\ref{t:run2}}\label{ap:proof1}

Notice that this game can be rephrased as follows. The oracle's first answer is always proper, i.e.: 
$g^{r}, g^{r \alpha}, g^{r \alpha_{i_1}}, \dots g^{r \alpha_{i_{t-1}}}$. 
\iffalse
{\footnotesize
\[\begin{array}{ccccc}
g^{r}, & g^{r \alpha}, & g^{r \alpha_{i_1}}, & \dots & g^{r \alpha_{i_{t-1}}}\\
\end{array}\]}%
\fi
Only the following answers are constructed either always properly (if $b=0$), or always randomly. 
It follows from the fact that $t$ random values (in the first oracle's answer) always define a polynomial of degree at most $t-1$.

\begin{proof}[sketch]
Assume that $A$ asks the oracle for partial decryptions at most $d$ times (where $d$ is polynomial in $k$).
For simplicity, we assume that $n=t=3$ and $d=2$. The proof for greater $n$, $t$, and $d$ can be made similarly.

Assume to the contrary that $A$ winning the game with non-negligible advantage $a$, exists. 
Using $A$ we construct an adversary $A^*$ that breaks the ElGamal security. 
Firstly, $A^*$ receives a public key $(q,g,g^\alpha)$ from a ``semantic security'' oracle. 
%, and forwards it to $A$.
Secondly, $A^*$ generates $x_1, x_2, x_3 \in_R \mathbb{Z}_q$ and sends them to $A$. 
Then $A^*$ sends plaintexts $p_0\text{=}1$ and $p_1 \in_R \mathbb{G}$ to the oracle, which answers with $g^{r_1}, p_b g^{r_1 \alpha}$.

Now, $A^*$ chooses a random permutation $\pi:\{1,2,3\} \rightarrow \{1,2,3\}$ (we denote $j_f=\pi(f)$), and
picks $\alpha_{j_1}, \alpha_{j_2} \in_R \mathbb{Z}_q$. Then $A^*$ computes (using Equation~\ref{eq:int}),
such $g^{\alpha_{j_3}}$ that points: 
%{\footnotesize\[
$\{(0,\alpha_{j_1}), (x_{j_1},\alpha), (x_{j_2},\alpha_{j_2}), (x_{j_3},\alpha_{j_3}) \}$
%\]}%
define a polynomial of degree $2$.
We denote (for $1 \le i \le 3$): $\alpha'_i = \alpha$ if $i=j$, and $\alpha'_i = \alpha_i$ otherwise. 
$A^*$ sends a public key $pk'=(q,g,g^{\alpha_{j_1}})$ to $A$. 

When $A$ asks the first time (for partial decryptions) with $i_1, i_2$ then $A^*$ answers (for $r \in_R \mathbb{Z}_q$) with: 
$g^{r}, g^{r \alpha_j}, g^{r \alpha'_{i_1}}, g^{r \alpha'_{i_2}}$. 
\iffalse
{\footnotesize
\[\begin{array}{cccc}
g^{r}, & g^{r \alpha_j}, & g^{r \alpha'_{i_1}}, & g^{r \alpha'_{i_2}}\\
\end{array}\]}%
\fi
For the second $A$'s question $i'_1,i'_2$, $A^*$ firstly checks whether $\{i'_1,i'_2\} \not= \{j_1,j_2\}$. 
If it holds then $A^*$ halts and outputs a random bit.
Otherwise 
%$A^*$ sends the message to $A$ gradually. 
$A^*$ first sends $g^{r_1}, g^{r_1 \alpha_j}$.
Then $A^*$ chooses $b' \in_R \{0,1\}$, and for every $1 \le e \le 2$, acts as follows. 
If $i'_e=j_1$ then $A^*$ sends $p_b g^{r_1 \alpha}$ to $A$.
If $i'_e = j_2$ and $b'=0$ then $A^*$ sends $g^{r_1 \alpha_{j_2}}$.
Otherwise ($i'_e = j_2$ and $b'=1$): $g^{r_1 x}$ (for $x \in_R \mathbb{Z}_q$) is sent. 
Finally, $A^*$ returns the $A$'s output.

Notice that the probability that $\{i'_1,i'_2\}\not=\{j_1,j_2\}$ (and that $A^*$ halts with a random output) is $1 - 1/\binom{3}{2}$. 
Assume that it does not happen. 
If $b=b'$ then $A$'s input is well constructed and the probability that $A$ outputs $b$ is $\frac{1}{2} + a$.
Otherwise, because of the random permutation $\pi$, $A$'s input is distributed independently of $b$. 
Hence, the probability of $A$ guessing correctly is $\frac{1}{2}$ in this case. 
Therefore, $A^*$'s advantage is $a / (2 \binom{3}{2})$, and is non-negligible.
\qed
\end{proof}

The full proof for this lemma is similar, but complex, and we omit it here due to the space constraints 
(the proof for $d>2$ uses similar techniques as in the proof of Lemma~\ref{t:run1}).

\section{Proof Sketch of Lemma~\ref{theorem:privacy_server}}\label{ap:proof2}

\begin{proof}[sketch]
%In order to prove the theorem we construct a simulator $I$ that works in the ideal situation and acts as a server to an adversary $A$ working in the real situation. 
%We ensure that data sent by $I$ to $A$ are computationally indistinguishable from the data send by the server (Figure~\ref{scheme2}).
We construct $I$ that works only for unsuccessful PR invocations. The proof for a successful $A$'s invocation can be made similarly. 

Firstly, $I$ generates $v_1 \in_R \{0,1\}^k$, a public key $pk=(q, g, g^\alpha)$, $y \in_R \mathbb{D}^n$, 
and shares $(x_1, \alpha_1), \dots (x_n, \alpha_n) \in {\mathbb{Z}_q}^2$ of the equivocable TE scheme (Section~\ref{threshold}), such that $x_i=\mathfrak{h}_i(p_i)$.
Later, when $A$ sends his $i$th guess ${p}^i$, then $I$ forwards it to the ``ideal'' oracle. 
If the oracle's answer equals $p$ then $I$ halts. 
Otherwise $I$ chooses $r, r_1, \dots r_n \in_R \mathbb{Z}_q$, and sends: $v_1$, $pk$, $c$=$(g^r, y g^{r \alpha})$, $\{c_1$=$g^{r r_1}, \dots c_n$=$g^{r r_n}\}$ to $A$.

$A$ can only submit the proper guess of the password (otherwise the server would recognize it). 
Therefore, $A$ cannot break the protocol by disobeying the PR routine. 
Hence, now we only need to show that the $A$'s view send by $I$ is indistinguishable from the corresponding 
view in the real situation. 
%(Figure~\ref{scheme2}). 

Let's now consider $A$ in the real situation (Figure~\ref{scheme2}).
Notice that, because $A$ works in ROM, the data received by $A$ in any $d$ unsuccessful PR runs corresponds to the data from $d$ unsuccessful invocations of 
the algorithm $CM$ in the equivocable TE scheme. 
The difference is that, here, $A$ does not know which value $\mathfrak{h}_i(p_i)$ (for any $p_i \in \mathbb{D}$) is a part of a share (i.e., equals $x_i$), while $CM$ correctly knows all $x_i$. 
However, Lemmas~\ref{t:run1}, \ref{t:run2}, and Corollary~\ref{col1} can be applied in $A$'s case, because $A$ has actually less information than 
the combiner $CM$. 

In every invocation $A$ receives at most $t-1$ correct partial decryptions. 
Incorrect partial decryptions are created using values independent of $\alpha$ and $\alpha_i$, 
because if $p_j \not= p'_j$ then $\alpha_j - \mathfrak{g}_{j}(p_j) + \mathfrak{g}_{j}(p'_j)$ is random in $\mathbb{Z}_q$ (in ROM).
Therefore, based on Lemma~\ref{t:run1}, $A$ cannot recognize encryptions received in the real situation from encryptions received from $I$. 

Let $p$ be any password from $D^n$ encoded in $\mathbb{G}$, and ${p}^1, \dots {p}^d$ is any list of passwords not similar to $p$. 
Consider the following probability distributions of instances of the adversary's view:
\vspace*{-0.25cm}\begin{itemize}
\item $S_{0,0}$: 
$A$ receives properly constructed data from the the PR routine (Figure~\ref{scheme2}) for his guesses ${p}^1, \dots {p}^d$, and for the password $p$. 
\item $S_{0,1}$:
For every guess ${p}^i$, $A$ receives proper $v_1$, $pk$, an encryption of $p$: $c$, and $n$ values:
if ${p}_j^i=p_j$ ($1 \le j \le n$) then a correct partial decryption $c'_i$, and $c'_j \in_R \mathbb{G}$ otherwise. 
\item $S_{1,1}$:
similar to $S_{0,1}$, but all $c'_j \in_R \mathbb{G}$ (for every guess); $S_{1,1}$ corresponds to the view sent by $I$. 
%$A$ receives the data constructed as follows. For every his guess ${p}^i$ ($1 \le i \le d$): all $c'_j \in_R \mathbb{G}$.
\end{itemize}

%\looseness=-1
\vspace*{-0.25cm}We show that no algorithm $\mathfrak{D}(p,{p}^1, \dots {p}^d)$ 
can distinguish between an input sampled from $S_{0,0}$
and an input sampled from $S_{1,1}$ (under the DDH assumption). 
Define $\mathfrak{D}_{0,0}$  as the probability that the output of $\mathfrak{D}$ is $1$ given an input sampled from $S_{0,0}$. 
Similarly, we define $\mathfrak{D}_{1,1}$, $\mathfrak{D}_{0,1}$. 
It holds that 
\vspace*{-0.25cm}\[
|\mathfrak{D}_{0,0}-\mathfrak{D}_{1,1}| \le |\mathfrak{D}_{0,0}-\mathfrak{D}_{0,1}| + |\mathfrak{D}_{0,1}-\mathfrak{D}_{1,1}|. 
\vspace*{-0.25cm}\]
Assume to the contrary that $|\mathfrak{D}_{0,0}-\mathfrak{D}_{1,1}|$ is non-negligible. 
Then, either $|\mathfrak{D}_{0,0}-\mathfrak{D}_{0,1}|$ is non-negligible or $|\mathfrak{D}_{0,1}-\mathfrak{D}_{1,1}|$ is non-negligible. 
In the first case, Corollary~\ref{col1} does not hold. 
%, while 
In the second case, Lemma~\ref{t:run2} does not hold. 
Therefore, $A$ cannot distinguish $I$ from the real server under the DDH assumption. 
%'s view from unsuccessful runs is indistinguishable from a view prepared by $I$.
\qed
\end{proof}

\end{document}